# Family of Integrated Multi-Input Multi-Output DC-DC Power Converters

Bang Le-Huy Nguyen[1*], Honnyong Cha[1], Tien-The Nguyen[1] and Heung-Geun Kim[2]
[1] School of Energy Engineering, Kyungpook National University, Daegu, Korea
[2] Department of Electrical Engineering, Kyungpook National University, Daegu, Korea
*E-mail: bangnguyen@ieee.org

*Abstract*— This paper explores a family of integrated multiport converters using three-switch which can provide single-input dual-output (SIDO) or dual-input single-output (DISO) with bidirectional power flow between any two ports. The concept can be extended to the *n*-switch converters to achieve more inputs and/or outputs. The proposed converters can be applied to interfacing sources, loads and storage elements having different voltage levels in applications such as dc nanogrids, electric vehicle, multiport power supplies, distributed generation systems. Various topological configurations of the integrated multiport *n*-switch converter are investigated. The operating principles and PWM control strategy of these converters are analyzed in detail. A universalized hardware prototype is built, experimental results are provided for verification.

## I. Introduction

The dc-dc power conversion plays a major role in the power electronics field spreading from low to high power range. Recently, the development of distributed generation systems, dc nanogrids, and electric vehicle, which are included many energy sources, loads, and storage elements, leads to the strong demand for compact integrated converter systems. Conventionally, to supply to different loads, to obtain energy from different sources, and to provide bidirectional power transfer for storage components, discrete power converters are implemented and regulated separately. These converters are coordinated to manage the power flow between them through a common control system via communication channels.

For more effectively monitor and cooperative control these converter systems, the concept of multiport power electronic interface (MPEI) [1]–[3] was introduced and implemented. This architecture considers all related converters as a united converter with a dynamic model [1] to enhance the transient stability and reliability. Fig. 1(a) shows the MPEI using separated dc-dc converters. For further improvements, the integrated multiport dc-dc converters as in Fig. 1(b) appears to reduce the number of redundant components and to make the system more compact and lower cost.

The integrated multiport converter topologies are more and more attractive with many topologies introduced in

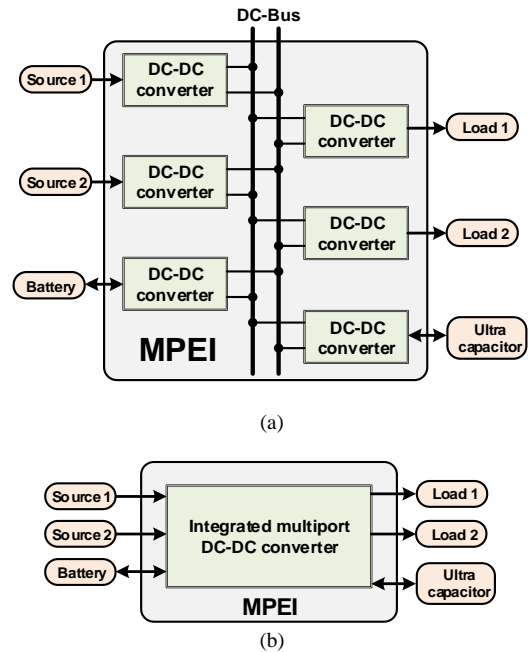

Fig. 1. Multiport power electronics interface architecture with separated dc-dc converters (a) and integrated multiport converter (b).

the literature. The paper [4]–[6] presented two families of multi-input single-output (MISO) converter topologies by combining some pulsating source cells with one output filter. Whereas, in [7], one pulsating source cell was combined with different filter cells to make single-input multi-output (SIMO) converter topologies. In [8], The family of three ports converters is introduced. However, some of these converters cannot provide bidirectional power flow which is crucial for systems involving energy storage elements. Besides, the complexity and inflexibility in structure make these converters ineffective. In [9], the multi-input bidirectional converter is introduced to interface a battery, an ultra-capacitor, and the dc-link of the hybrid electric vehicle/ fuel cell vehicle (HEV/FCV) application. This converter is derived by combining two non-inverting buck/boost converter with a sharing leg to provide both buck and boost function between any two ports. However, it uses more switches and the current in the battery is discontinuous. A bidirectional SIDO buck converter using only three switches were proposed in [10]–[12]. Similarly, the paper

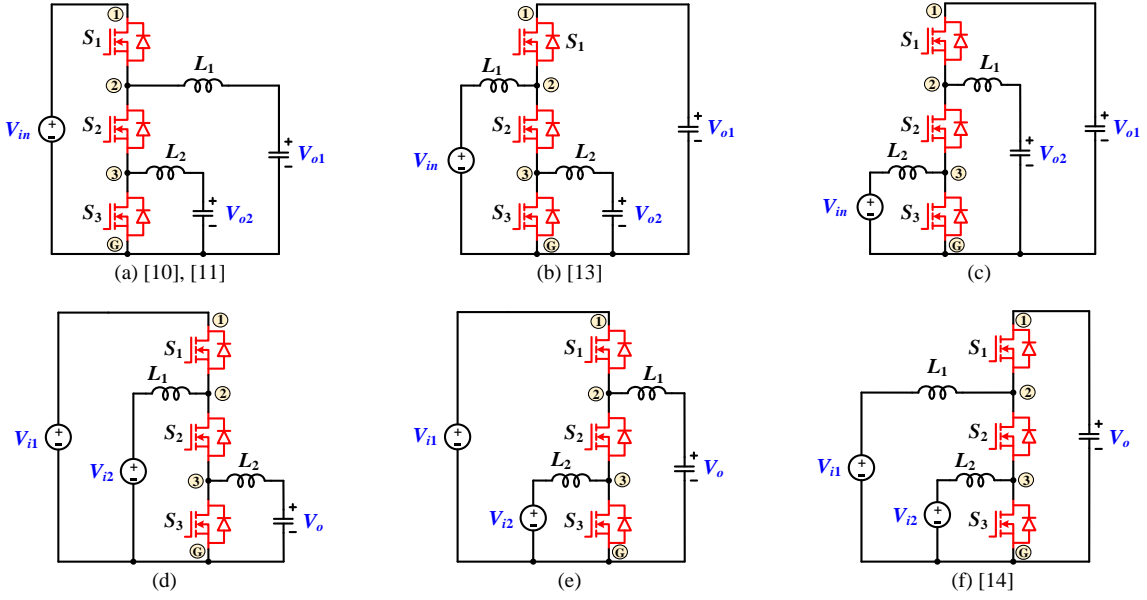

Fig. 2. Topological configurations of three-switch converter for SIDO and DISO converters. (a) two buck outputs [10], [11] (b) one boost and one buck outputs [13], (c) two boost outputs, (d) two buck inputs, (e) one buck and one boost inputs, and (f) two boost inputs [14].

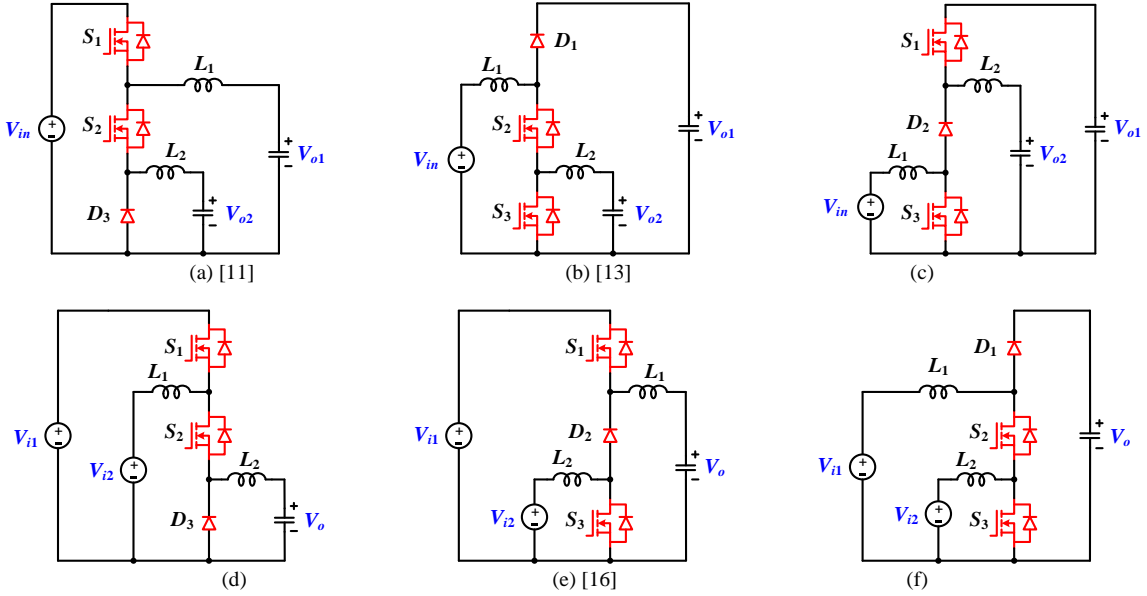

Fig. 3. Nonsynchronous derivations of SIDO and DISO converters. (a) two buck outputs [11], (b) one boost and one buck outputs [13], (c) two boost outputs, (d) two buck inputs, (e) one buck and one boost inputs [16], and (f) two boost inputs.

[13] introduced another configuration of three-switch to achieve one buck and one boost outputs. These converters are also extended to more switches to achieve more buck outputs. In [14]–[15], a DISO using three-switch was introduced. More generally, in [16], a topological synthesis method is proposed to derive nonsynchronous SIDO and DISO three-switch converters from conventional SISO converters. These above converters have benefits of the minimum number of switches and the integration in both hardware and control system.

Interestingly, other SIDO and DISO converters using three switches are still veiled in these papers [10]–[17]. This paper explores more topological configurations of three-switch converters. Gathering with the existed three-switch converters, now the whole family of the integrated multiport converter which can provide bidirectional power between three ports is fully explored. The nonsynchronous versions of this converter family are also derived. Intensively, the principle and synthesis method for various multiport $n$-switch converter configurations are proposed and investigated. The operating principle and PWM control strategy are also analyzed in detail.

## II. TOPOLOGICAL CONFIGURATIONS

In this section, firstly, the bidirectional three-switch SIDO and DISO converters with the common-ground ports are described. Secondly, the nonsynchronous derivations of these converters are considered. The extension to the $n$-switch converter is also discussed.

Then, other derivations of three-switch converters with different-ground ports are investigated. Finally, the universalized configurations which are the multi-purposes converters are proposed and analyzed.

*A. Common-Ground Topologies*

By adding a switch to the conventional buck converter to have one more output, the first SIDO converter using three-switch were proposed in 2006 [10] as shown in Fig. 2(a). In [11], this converter topology is analyzed in detail, its nonsynchronous version shown in Fig. 3(a) is also introduced. By adding a switch to conventional boost converter to get one more buck output, other SIDO converter is proposed in [13] as shown in Fig. 2(b). In [14], by combining two boost converters, a DISO converter is proposed as shown in Fig. 2(f). In the paper [16], the basic source and load cells are defined. Inserting these cells to the basic dc-dc converter topologies (buck, boost, buck-boost, Cuk, sepic, zeta), six SIDO and six DISO converters are proposed included those shown in Figs.2 (a) and (e).

More simply, there are four nodes in the three-switch converter as denoted in Figs. (a)–(f). Except for the ground node G, any other nodes can be connected to the input source or output load. When one node is connected to the input and the remained nodes are the outputs, the three-switch SIDO converter family can be achieved as in Figs. 2(a)–(c). The converter shown in Fig. 2(c) is a new member with two boost outputs. On the other hand, if two nodes are used for the inputs and the remained node is connected to the output, the three-switch DISO converter family is created as in Figs. 2(d)–(f).

*B. Nonsynchronous Derivations*

The three-switch converters in Figs. 2(a)–(f) have the bidirectional power transfer between any two ports. However, if only unidirectional power flow is required, the nonsynchronous derivations are considered to reduce the number of active switches and accompanied circuits as gate driver power supply and gate driver circuit. Moreover, there is no shoot-through path in the nonsynchronous converters, then the dead-time protection can be removed.

The nonsynchronous derivations of the above three-switch converter can be achieved from the synchronous ones by replacing the switch with a diode as shown in Figs. 3(a)–(f), respectively. Where the switches, which are possible to be replaced, always conduct current in its diode.

Summarily, the family of SIDO and DISO dc-dc converters with common-ground ports using three-switch included the nonsynchronous derivations is fully explored.

*C. n-Switch Extension*

Extensively, the three-switch converters can be extended to *n*-switch as shown in Fig. 4. A similar approach of synthesis principle can be applied. Except for the ground node G, the remained *n* nodes of this *n*-switch converter can be connected to inputs or outputs. Thus, any *k*-input and *p*-output converters can be derived, where $n = (k+p)$. Entirely, the total number of different multiport converters are ($2^n - 2$) as described in Table I. Notably, all those multiport converters have the same ground of inputs and outputs.

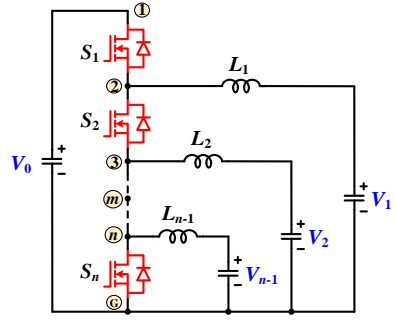

Fig. 4. The common-ground multiport *n*-switch converter.

TABLE I
THE NUMBER OF *N*-SWITCH CONVERTERS

| Number of input ports | Number of output ports | Number of converters |
|---|---|---|
| 1 | $n-1$ | $C_n^1$ |
| 2 | $n-2$ | $C_n^2$ |
| ... | .... | .... |
| $n-1$ | 1 | $C_n^{n-1}$ |
| $C_n^1 + C_n^2 + ... + C_n^{n-1} = 2^n - 2$ | | |

*D. Different-Ground Topologies*

The paper [17] introduced a SIDO using three-switch with different-ground outputs as shown in Fig. 5(a). Where $V_0$ is connected to the input, whereas $V_1$ and $V_2$ are connected to outputs. Reversely, using $V_1$ and $V_2$ as inputs and $V_0$ as output, the paper [15] formed a DISO converter. Where the port $V_1$ is connected to nodes 2 and 3. Whereas the port $V_2$ is connected to node 3 and ground (G).

Seemingly, except for the fixed node $V_0$, the ports $V_1$ and $V_2$ can be connected to other two nodes. More topologies, therefore, can be achieved as described in Figs. 5(b)–(g). In Figs. 5(a)–(c), the ports are connected to the adjacent nodes, whereas the further-distance nodes are combined in Figs. 5(d)–(g).

Similar to the analysis of the common-ground multiport converter topology, each topology shown in Fig. 5 also possess abilities as follows:

1) They can form three SIDO and three DISO converters by arranging the inputs and outputs by the same way shown in Fig. 2.

2) The nonsynchronous versions of these SIDO and DISO converters can also be derived.

3) They can be extended to the *n*-switch converters.

Considerably, when extending to *n*-switch, the number of nodes becomes ($n+1$). Beside the port $V_0$ is connected to nodes 1 and G, other ($n-1$) ports can be connected to any two nodes. A myriad of topological configurations can be formed. As a result, various integrated MIMO dc-dc power converters can be achieved.

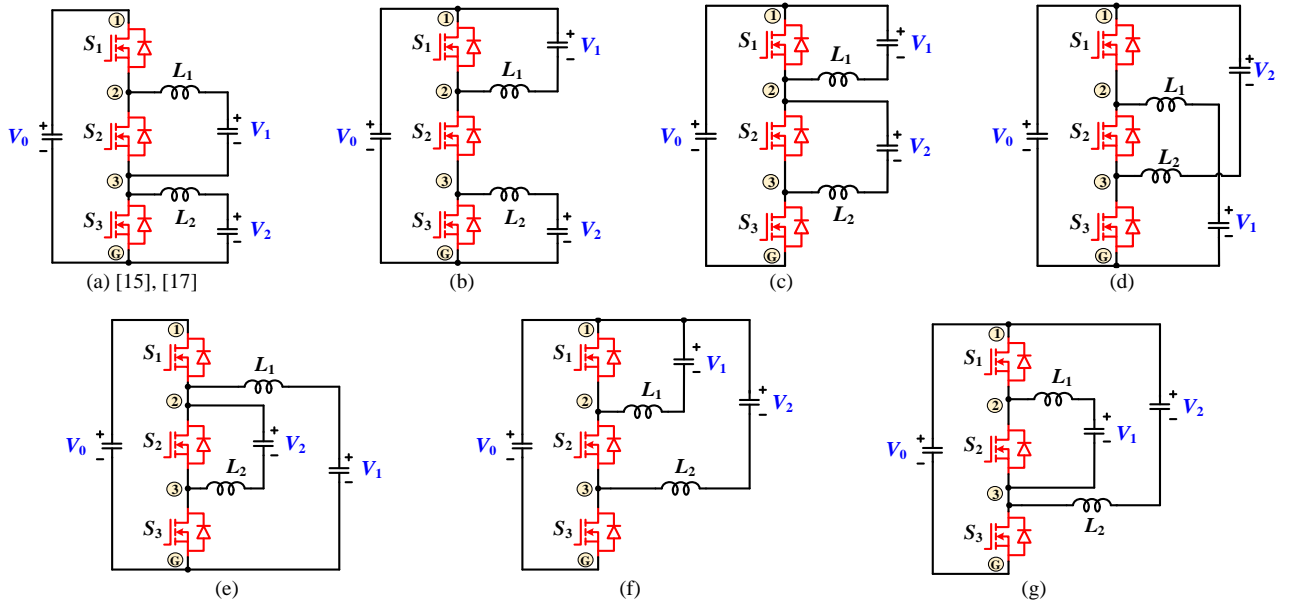

Fig. 5. Topological configurations of the three-switch converters with different-ground ports.

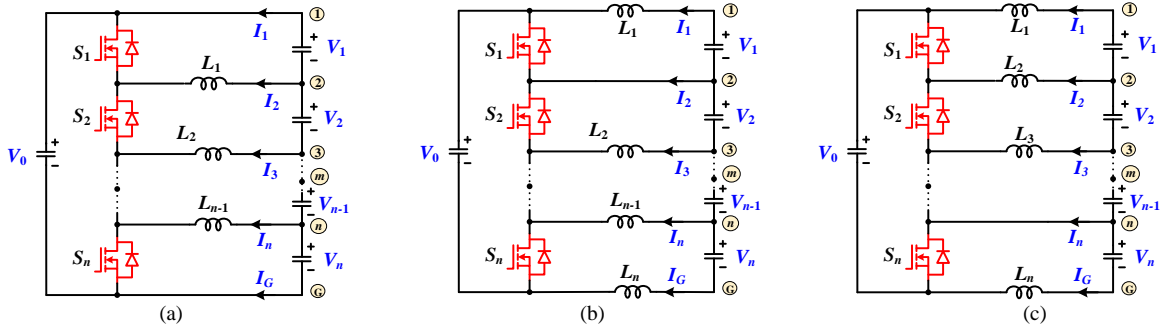

Fig. 6. The multi-purpose integrated $n$-switch multiport converters.

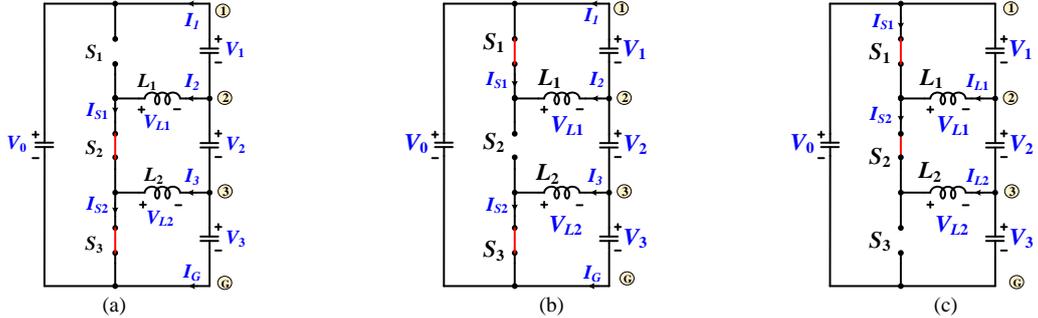

Fig. 7. The operating modes of the multi-purpose three-switch converter (a) Mode 1. (b) Mode 2. (c) Mode 3.

*E. Universalized Configurations*

The universalized configurations of the integrated multiport converter are derived as shown in Fig. 6. They are the multi-purpose $n$-switch converters which can replace equally all above topologies. Where the pulsating voltage across each switch is filtered to the port by the *LC* filter. All ports are connected in series and the nodes now are placed between the ports. Hence, the input sources and/or output loads can feed and draw the power by connecting to any two nodes.

The $n$-switch multiport converter now becomes an energy router which diverts and allocates the power from and to its ($n$+1) ports. The total voltages of ports $V_1$–$V_n$ is the port voltage $V_0$. The port voltages can be controlled by the duty ratios of the responding switches, respectively, as in equation (9) ($V_i = D_i V_0$, $i$: 1–$n$) which is derived in Section III. Where the duty ratios are defined as the proportion of the period when turning off the switch over the total duty cycle time.

Fig. 6(a) shows the configuration in which the currents $I_2 - I_n$ are continuous owing to the presence of inductors. Whereas the currents $I_1$ and $I_G$ are discontinuous. In some applications, if the currents $I_1$ and $I_G$ are required to be continuous. The inductors can be placed in their paths as shown in Figs. 6(b)–(c). Where one inductor in any middle branches can be removed to reduce the number of inductors.

TABLE II
CURRENT FLOWING THROUGH SWITCHES

| Mode | $S_1$ | $S_2$ | $S_3$ | $S_4$ | … | $S_n$ |
|---|---|---|---|---|---|---|
| 1 | **0** | $I_2$ | $\Sigma I_{2\text{-}3}$ | $\Sigma I_{2\text{-}4}$ | … | $\Sigma I_{2\text{-}n}$ |
| 2 | $-I_2$ | **0** | $I_3$ | $\Sigma I_{3\text{-}4}$ | … | $\Sigma I_{3\text{-}n}$ |
| 3 | $-\Sigma I_{2\text{-}3}$ | $-I_3$ | **0** | $I_4$ | … | $\Sigma I_{4\text{-}n}$ |
| … | … | … | … | … | … | … |
| n-1 | $-\Sigma I_{2\text{-}(n-1)}$ | $-\Sigma I_{3\text{-}(n-1)}$ | $-\Sigma I_{4\text{-}(n-1)}$ | $-\Sigma I_{5\text{-}(n-1)}$ | … | $I_n$ |
| n | $-\Sigma I_{2\text{-}n}$ | $-\Sigma I_{3\text{-}(n-1)}$ | $-\Sigma I_{4\text{-}(n-1)}$ | $-\Sigma I_{5\text{-}(n-1)}$ | … | **0** |

## III. STEADY-STATE ANALYSIS AND PWM CONTROL STRATEGY

For simplicity, the port voltages are assumed to be stable at a fixed value in steady-state and the inductor currents are continuous. In the operation of these multiport converters, only one switch is turned off at the time, whereas all remained switches are turned on. Thus, there are $n$ modes for the $n$-switch converters. Figs. 7(a)–(c) show the equivalent circuits in the operating modes of the integrated multi-purpose three-switch multiport converter. The analysis of three modes is discussed below.

*1) Mode* 1 (duty ratio $D_1$)*:* the switch $S_1$ is off, while both switch $S_2$ and $S_3$ are on. The voltage imposing on inductors $L_1$, $L_2$ are as follows:

$$V_{L1} = V_1 - V_0 \; , \; V_{L2} = -V_3 \quad (1)$$

*2) Mode* 2 (duty ratio $D_2$)*:* the switch $S_2$ is off, while both switch $S_1$ and $S_3$ are on. The voltage imposing on inductors $L_1$, $L_2$ are as follows:

$$V_{L1} = V_1 \; , \; V_{L2} = -V_3 \quad (2)$$

*3) Mode* 3 (duty ratio $D_3$)*:* the switch $S_3$ is off, while both switch $S_1$ and $S_2$ are on. The voltage imposing on inductors $L_1$, $L_2$ are as follows:

$$V_{L1} = V_1 \; , \; V_{L2} = V_0 - V_3 \quad (3)$$

Apply the flux (or volt-sec) balance condition for inductor $L_1$ and $L_2$, respectively, we have:

$$D_1(V_1 - V_0) + (D_2 + D_3)V_1 = 0 \quad (4)$$

$$-(D_1 + D_2)V_3 + D_3(V_0 - V_3) = 0 \quad (5)$$

Where,

$$D_1 + D_2 + D_3 = 1 \quad (6)$$

$$V_1 + V_2 + V_3 = V_0 \quad (7)$$

According to equations (4)–(7), the gains between the port voltages can be determined as:

$$V_1 = D_1 V_0 \; ; \; V_2 = D_2 V_0 \; ; \; V_3 = D_3 V_0 \quad (8)$$

Extensively, the port voltage $V_i$ can be determined by the duty ratio $D_i$ ($i$: 1–$n$) as follows.

$$V_i = D_i V_0 \; , \; i\text{: }1\text{–}n \quad (9)$$

Using the duty ratios $D_1$–$D_n$ of switches $S_1$–$S_n$, all port voltages and inductor current can be regulated independently. The general gate drive signals and duty ratio control diagram are given as illustrated in Figs. 8 and 9, respectively.

The voltage stresses on switches are the voltage across nodes 1 and G ($V_0$). The currents flowing through the switches in the operating modes are described as in Table II. Where all inductor currents have the direction as denoted in Fig. 6. Apparently, the total current stresses on

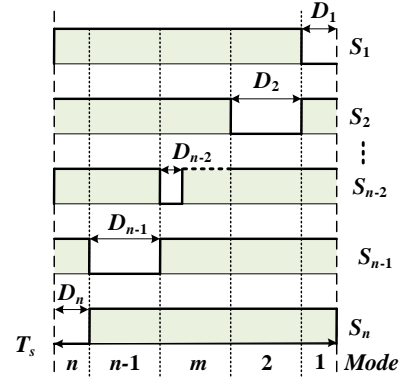
Fig. 8. Gate drive signals

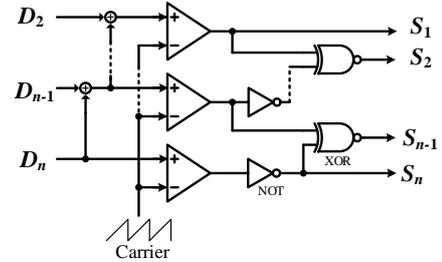
Fig. 9. The duty ratio control diagram

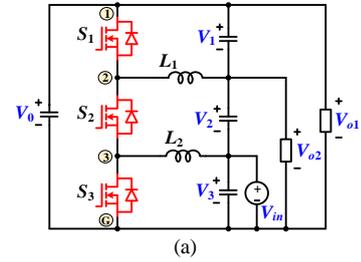
(a)

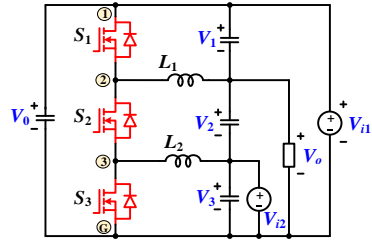
(b)
Fig. 10. The generalized three-switch multiport converter with sources and loads connected as SIDO in test 1 (a) and DISO in test 2 (b).

switches will be reduced provided that the inductor current is different in direction or the middle nodes (except for nodes 1 and G) are connected to inputs and outputs at the same time.

## IV. EXPERIMENTAL VERIFICATION

The universalized hardware prototype of the generalized three-switch multiport converter using MOSFETs 47N60CFD of Infineon is built and tested. The inductors are of 0.72 mH inductance value and the capacitors are of 560 µF capacitance value. The switching frequency is 30 kHz.

Firstly, one source and two loads are connected to form a SIDO converter as shown in Fig. 10(a). Where, the input voltage $V_{in} = 40$ V and the duty ratios $D_1 = 0.35$,

$D_2 = 0.25$, and $D_3 = 0.4$. Two outputs are the resistive loads of 50 Ω. The experimental results of this test are given in Fig. 11. The output voltages are calculated as $V_{o1} = V_3/D_3 = 40/0.4 = 100$ V; $V_{o2} = V_2+V_3 = D_2V_3/D_3 +V_3 = 65$ V. These calculated values are verified by the voltage waveforms of $V_{o1}$ and $V_{o2}$ shown in Fig. 11. The load currents are the average values of the inductor currents as the $I_{L1}$ and $I_{L2}$ waveforms.

In the second test, two sources and one load are used to achieve a DISO converter as shown in Fig. 10(b). Where the input voltages are $V_{i1} = 50$ V, $V_{i2} = 20$ V and the duty ratios are the same as the first test. The output is the resistive load of 50 Ω. Where, the output voltage $V_o$ can be calculated as $V_{o1}= V_2+V_3 = D_2V_{i1}+V_3 = 32.5$ V. The experimental inductor currents and output voltage waveforms shown in Fig. 12 are agreed with the theoretical analysis. The output load and input source currents of $V_{i2}$ are the average value of the inductor currents $I_{L1}$ and $I_{L2}$ waveforms, respectively.

## V. CONCLUSIONS

In this paper, the topological configurations for SIDO and DISO three-switch converters with both common and different ground of ports are explored and analysis in detail. The extension to *n*-switch is explained. The multi-purpose topologies which can be employed as an energy router are proposed. The steady-state analysis and PWM control strategy for the *n*-switch converters are provided. The gate signals and duty ratio control diagram are given. The hardware prototype is built and tested, the experimental results are agreed with the theoretical analysis.

ACKNOWLEDGMENT

This work was supported by the Korea Institute of Energy Technology Evaluation and Planning (KETEP) and the Ministry of Trade, Industry & Energy (MOTIE) of the Republic of Korea (No. 20174030201490) and was also supported by Basic Science Research Program through the National Research Foundation of Korea (NRF) funded by the Ministry of Education (NRF-2016R1D1A1B03934577).

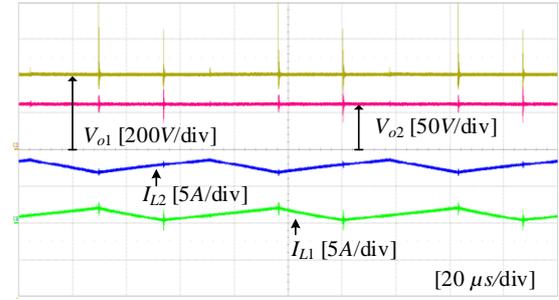

Fig. 11. The experimental waveforms for the SIDO converter in Fig. 10(a) with $D_1 = 0.35$, $D_2 = 0.25$, and $D_3 = 0.4$.

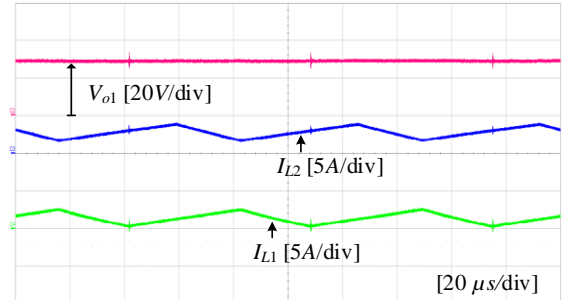

Fig. 12. The experimental waveforms for the DISO converter in Fig. 10(b) with $D_1 = 0.35$, $D_2 = 0.25$, and $D_3 = 0.4$.